\documentclass[11pt]{article}
\usepackage{moriond,epsfig}

\def\be{\begin{equation}}
\def\ee{\end{equation}}
\def\bea{\begin{eqnarray}}
\def\eea{\end{eqnarray}}

\def\lsim{\raise0.3ex\hbox{$<$\kern-0.75em\raise-1.1ex\hbox{$\sim$}}}
\def\gsim{\raise0.3ex\hbox{$>$\kern-0.75em\raise-1.1ex\hbox{$\sim$}}}

\def\pom{{I\!\!P}}

\def\beq{\begin{equation}}
\def\eeq{\end{equation}}
\def\bea{\begin{eqnarray}}
\def\eea{\end{eqnarray}}
\def\bq{\begin{quote}}
\def\eq{\end{quote}}

\begin{document}
\vspace*{4cm}
\title{ASPECTS OF THE UNITARIZED SOFT MULTIPOMERON APPROACH IN DIS AND
DIFFRACTION}

\author{\underline{E. G. FERREIRO}$^{(a)}$, M. B. GAY DUCATI$^{(b)}$,  
M. MACHADO$^{(b)}$, C. SALGADO$^{(c)}$ }

\address{$^{(a)}$ Depto. de F\'{\i}sica de Part\'{\i}culas,
Universidade de Santiago de Compostela \\
E-15706 Santiago de Compostela, Spain\\
$^{(b)}$ Instituto de F\'{\i}sica, Universidade Federal do Rio
Grande do Sul,\\ Caixa Postal 15051, CEP 91501-970, Porto Alegre, RS, Brazil\\
$^{(c)}$ CERN CH-1211 Geneva 23, Switzerland
}

\maketitle\abstracts{We study in detail the main features of the unitarized
Regge model (CFKS),
recently proposed to describe the  small-$Q^2$ domain. It takes into account
a two-component description
with two types of unitarized contributions:
one is the multiple Pomeron
exchanges contribution, interacting with the large dipole size
configurations, and the other one consists on a unitarized dipole cross
section, describing the
interaction with the small size dipoles.
We compare the resulting dipole cross section to that
from the saturation model (GBW).
}

\section{Introduction}
The study of a new regime of QCD, that of high density of partons, has
drawn much attention in the
last
years. The key discovery was the
observation at HERA of the fast growth of
parton densities (mainly gluons) as the energy increases in experiments of
deep inelastic scattering.
Taking $\sigma^{tot}\sim s^{\alpha(0)-1}$ ($F_2\sim x^{-\alpha(0)+1}$),
values of $\Delta\equiv \alpha(0)-1$ in the range $0.1$ -- $0.5$ have been
reported, depending on the virtuality $Q^2$ of the photon.
However, 
some kind of saturation of this growing due to unitarity
effects
is
expected,
leading to the expected
limit given by the  Froissart bound ($\sigma\ \lsim\  (\log s)^2$ as
$s\to\infty$)\cite{frois}.

Taking into account that the saturation phenomenon is required in a complete
understanding of the high energy reactions, and that a consistent treatment of
both inclusive and diffractive processes should be done,  in
this work we study derivative quantities using the Regge unitarized  CFKS model
\cite{CFKS1,CFKS2}. In this hybrid model, both soft (multiperipheral
Pomeron and reggeon exchanges) and hard (dipole picture) contributions are
properly unitarized in an eikonal way with triple pomeron
interaction also included. This approach describes
the transition region and can be used as initial condition for a QCD
evolution at high virtualities.
The extrapolation to the higher-$Q^2$ domain
is also  performed here,
checking the behaviour of the model without including QCD evolution.
We discuss the similarities and/or connections  with
the phenomenological saturation model \cite{GW}, stressing that a QCD
evolution is required for a correct description of higher $Q^2$ in the
inclusive case. For the diffractive case, such a procedure is not formally
required, since the non-perturbative sector is dominant in this case.

\section{The inclusive case}

We start by briefly reviewing the CFKS approach. It  interpolates between
low
and high virtualities $Q^2$, which are related to the dipole separation size,
$r$,  at the target rest frame,  considering a two-component
model \cite{CFKS1,CFKS2}. Considering the unifying picture of the
color dipoles, the separation into a large size (in \cite{CFKS2} it is
called $L$) and a small size (called $S$ in
\cite{CFKS2}) components of the $q\bar{q}$ pair is made in terms of the
transverse distance $r$ between $q$ and $\bar{q}$. The border value,
$r_0$, is treated as a free parameter - which turn out to be $r_0 \sim
0.2$~fm.
Hereafter we use the notation
{\it soft} for the large size configuration and {\it hard} for the small size
one.

The soft component considers multiple
Pomeron exchanges (and
reggeon $f$) implemented in a quasi-eikonal approach.
It also includes the resummation of triple Pomeron branchings (the so-called
fan diagrams).
The initial input is a
phenomenological Pomeron with fixed intercept
$\alpha_P(0) = 1 + \varepsilon_{\pom}=1.2$
(further changes are due to absorptive corrections). In the
impact parameter representation,  the $b$-space,  
the Regge parametrization for the  amplitude
of the soft Pomeron exchange
looks like:
\begin{eqnarray}
\chi^{\pom}(s,b,Q^2) \simeq \frac{C_{\pom}}{R(x,Q^2)}
\left(\frac{Q^2}{s_0 + Q^2}\right)^{\varepsilon_{\pom}}\,
x^{-\varepsilon_{\pom}} \exp [-b^2/R(x,Q^2)] \,. \end{eqnarray}
The resummation of the triple-Pomeron branches is encoded in the
denominator of the amplitude $\chi^{n\,\pom}$, i.e. the Born term in the
eikonal expansion. Moreover, the corrected amplitude is eikonalized in the
total cross section,
\begin{eqnarray}
\chi^{n\,\pom}(x,Q^2,b)& = &
\frac{\chi^{\pom}(x,Q^2,b)}{1+a\chi_3(x,Q^2,b)}\,,\\
\sigma^{n\,\pom}(x,Q^2,b) & \simeq & 1-\exp \left[
-
\chi^{n\,\pom}(x,Q^2,b)\,\right]\,, \end{eqnarray}
where the constant $a$ depends on the proton-Pomeron and the triple-Pomeron
couplings at zero momentum transfer ($t=0$).

The eikonalization
procedure modifies the growth of the total cross section from a steep
power-like behavior to a milder logarithmic increase. The total soft
contribution is
 obtained by integrating over the impact parameter,
\begin{eqnarray}
\sigma^{soft}(s,Q^2)=4\,\int d^2b\,\,\sigma^{soft}(s,Q^2,b)\,.
\end{eqnarray}

The hard component is  considered in  the color dipole picture of DIS.
The dipole cross section, modeling the interaction
between the $q\bar{q}$ pair and the proton, $\sigma^{dipole}(x,r)$,   is taken
from the eikonalization of the  expression above $\chi^{n\,\pom}(s,b,Q^2)$
already corrected  by triple-Pomeron branching  (the fan
diagrams contributions).
The corresponding cross section  is extracted by considering the
contributions coming from  distances between 0 and  $r_0=0.2$ fm,
whereas for $r>r_0$ the contributions are described by the
soft piece already discussed. In such small distances, perturbative QCD is
expected to work.  The total cross
section considering this dipole cross section is expressed as:
\begin{eqnarray} \sigma^{hard}_{tot}(x,Q^2)
& = & \int_0^{r_0} d^2 r \, \int_0^1 d \alpha \, |\Psi^{T,L}_{\gamma^*
q}(\alpha,r)|^2  \,   \sigma^{dipole}_{CFKS}(x,r) \,,  \\
\sigma^{dipole}_{CFKS}(x,r) & = & 4\,\int d^2b \,\,
\sigma^{n\,\pom}(x,Q^2,b,r)\,,\\
\sigma^{n\,\pom}(x,Q^2,b,r) & \simeq & 1-\exp [
-
\,r^2 \chi^{n\,\pom}(x,Q^2,b)\,]\,,
\end{eqnarray}
\noindent
where $T$ and $L$ correspond to transverse and longitudinal polarizations
of a
virtual photon, $\Psi^{T,L}_{\gamma^*
q}(\alpha,r)$ are the corresponding wave functions of the
$q\bar{q}$-pair.

The $r^2$ dependence is introduced in the Born term of the eikonal
expansion, presented  in the last expression above, in order to ensure the
correct behavior determined by
the color transparency.
Thus a factor $r^2$ has
been
introduced in the eikonal of eq. (7).

The weight of each contribution (soft and hard) in the total cross section
[and $F_2(x,Q^2)$] can be obtained, providing an analysis of the
role played by each piece of the model:
\begin{eqnarray}
R_{SOFT}(x,Q^2)=
\frac{
\sigma_{tot}^{soft}(x,Q^2)}{
[\sigma_{tot}^{soft}(x,Q^2)+\sigma_{tot}^{hard}(x,Q^2)]}\,. \end{eqnarray}

\begin{flushleft}
\begin{tabular}{cc}
\psfig{figure=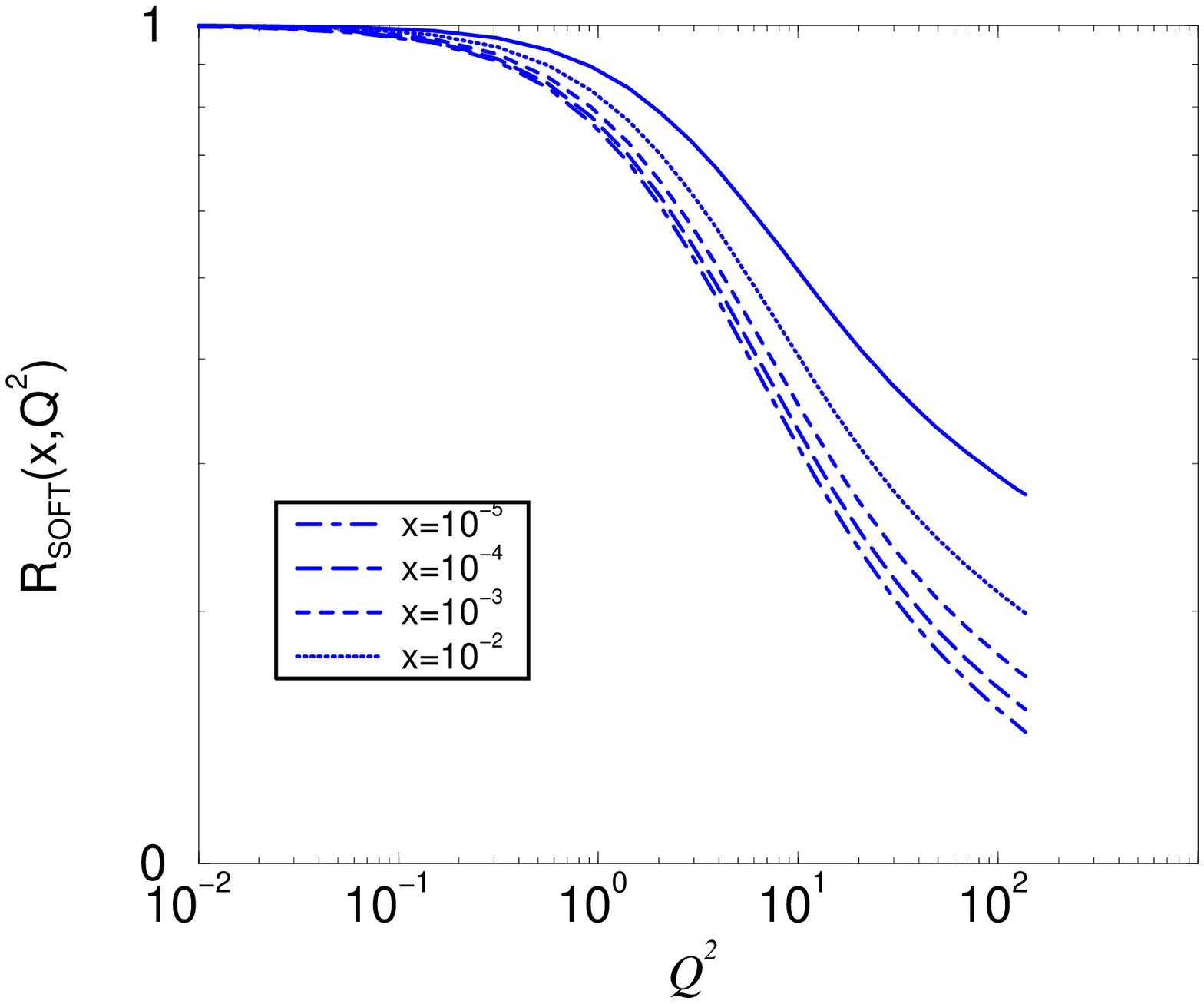,height=2.7in}
&
\psfig{figure=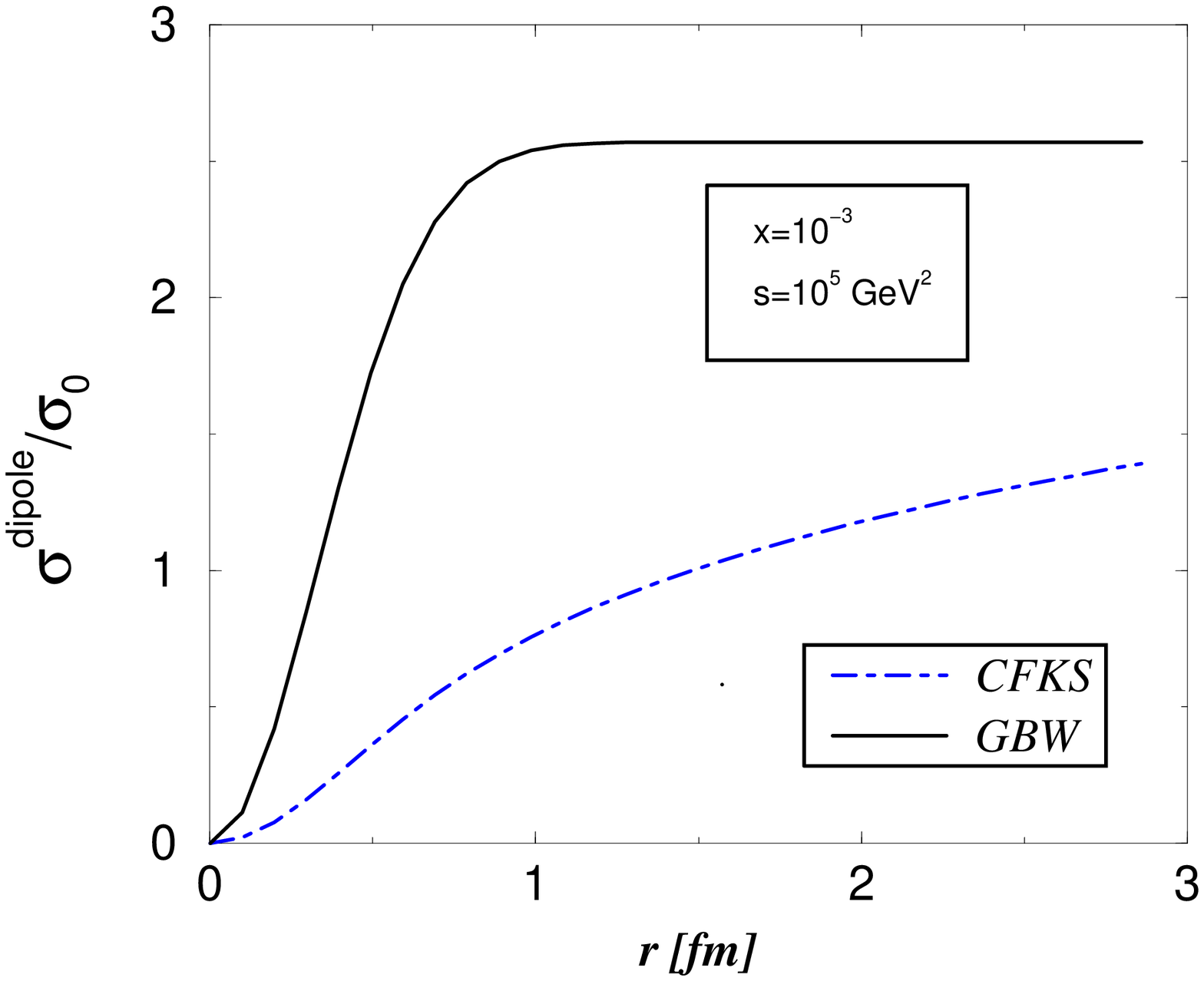,height=2.7in}\\
\begin{minipage}{7.7cm}
\vskip 10pt
\small{Figure 1. The ratio $R_{SOFT}$ as a function of $Q^2$ at fixed momentum
fraction $x$.}
\end{minipage}\hskip 0.5cm
&
\begin{minipage}{7.7cm}
\vskip 10pt
\small{Figure 2. The 
dipole cross section
from GBW and CFKS
as a function of the transverse dipole
separation $r$ at fixed $x$ ($s$).}
\end{minipage}
\end{tabular}
\end{flushleft}

Fig. 1 clearly shows that the soft piece is  dominant
at  $Q^2=0.01$ and decreases as the virtuality grows. The behavior is
monotonic, almost independent of the momentum fraction $x$.

An interesting issue  is the relation between the dipole cross section
coming from the CFKS model and the phenomenological one of
G.-Biernat-W\"{u}sthoff \cite{GW}. The GBW cross section is parametrized as:
\begin{eqnarray}
\sigma^{GBW} (x,r) & = & \sigma_0 \left[ 1- \exp(-r^2/4R^2_0(x))\right]\,,\\
R^2_0(x) & = & \left(\frac{x}{x_0}\right)^\lambda \rm{GeV}^{-2}\,,
\end{eqnarray}
where $\sigma_0=23.03$ mb properly normalizes the dipole cross section.  The
remaining parameters are $\lambda=0.288$ and $x_0=3.04\times 10^{-4}$,
all of them determined from the small-$x$ HERA data.  The $R_0(x)$ is
the main theoretical contribution, defining the saturation scale, which is
related with the taming of the gluon distribution at small $x$ (unitarity
effects) \cite{satmodels}. The above expression has been used to describe both
inclusive and diffractive structure functions, in good agreement with the
experimental results. The comparison between this approach and the
CFKS dipole cross section is shown in Fig. 2. The main feature of the GBW
parametrization is that it ensures that
the
dipole cross section  grows linearly with $r^2$ at small transverse separation,
whereas it saturates at large size configurations. The picture emerging from
the CFKS is slightly different,  presenting a mild (logarithmic) increase with
$r$, away from huge separation sizes that shifts the saturation scale up to
very high virtualities. Although the continuous and smooth increasing with
the radius, in the CFKS approach the cross section  underestimates the GBW one
for all $r$ --one should take into account that in the CFKS model there is also
a soft contribution--.

\section{The diffractive case}

The diffractive sector in the CFKS approach is constructed by a
three-component
model, using the AGK cutting rules to relate the
elastic multiple scattering amplitude to the inelastic diffractive
contribution. The first term comes directly from the soft piece, the
second one from the triple-Pomeron (and the reggeon $f$) interaction and the
last one from the hard (dipole) piece. The spectrum on $\beta$ is introduced by
hand, based on earlier
soft and hard (pQCD) calculations.

The agreement of CFKS approach with data is remarkable
even at high virtualities. In the saturation model \cite{GW}, 
the reliability of the
pQCD calculation is extended to smaller virtualities through the saturation
scale $R_0(x)$.

As a final study, we have performed 
the calculation of the $Q^2$ logarithmic slope of
the diffractive structure function $F_2^{D(3)}$.
 The motivation is that this
observable is a potential quantity to distinguish soft and hard dynamics
in diffractive DIS \cite{slope2}.
The
saturation model produces a transition between positive and negative slope
values at low $\beta=0.04$ (upper plots), while it shows a positive slope for
medium and large $\beta$. Instead, the CFKS approach  presents a positive
slope for the whole $Q^2$ and $x_{\pom}$ ranges, flattening at large $\beta$,
similarly as the non-saturated pQCD calculations \cite{slope2}.

\section{Conclusions}
A deeper understanding of the saturation phenomenon is required to perform
reliable estimations for the  current and forthcoming  high energy reactions.
The saturation scale, which sets the onset of the unitarity corrections, is
found to be in the transition regime of low $x$ and $Q^2$. In this
domain, both Regge-inspired phenomenology and improved pQCD calculations
(perturbative shadowing, higher twist),
considering unitarity effects, are able
to describe the data with high precision.
The most advantageous ones are those
describing in an unified way the inclusive processes as well as diffractive
ones.

\end{document}